\documentclass[10pt]{iopart}

\usepackage{graphicx} 
\graphicspath{{Images/}}

\usepackage{color}

\begin{document}
\title[Machine learning approach to muon spectroscopy analysis]{Machine learning approach to muon spectroscopy analysis
}
\author{T Tula$^1$, G M\"oller$^1$, J Quintanilla$^1$, S R Giblin$^2$, A D Hillier$^3$, E E McCabe$^1$, S Ramos$^1$, D S Barker$^{1,4}$, S Gibson$^1$ }
\address{$^1$ School of Physical Sciences, University of Kent,
Park Wood Rd, Canterbury CT2~7NH, United Kingdom}
\address{$^2$ School of Physics and Astronomy, Cardiff University, Cardiff CF24 3AA, United Kingdom}
\address{$^3$ 
ISIS Facility, STFC Rutherford Appleton Laboratory, Chilton, Didcot Oxon, OX11 0QX, United Kingdom}
\address{$^4$ School of Physics and Astronomy, University of Leeds, Leeds, LS2 9JT, United Kingdom}
\ead{J.Quintanilla@kent.ac.uk}
\begin{abstract}
	In recent years, artificial intelligence techniques have proved to be very successful when applied to problems in physical sciences. Here we apply an unsupervised machine learning (ML) algorithm called principal component analysis (PCA) as a tool to analyse the data from muon spectroscopy experiments. Specifically, we apply the ML technique to detect phase transitions in various materials. The measured quantity in muon spectroscopy is an asymmetry function, which may hold information about the distribution of the intrinsic magnetic field in combination with the dynamics of the sample. Sharp changes of shape of asymmetry functions -- measured at different temperatures -- might indicate a phase transition. Existing methods of processing the muon spectroscopy data are based on regression analysis, but choosing the right fitting function requires knowledge about the underlying physics of the probed material. Conversely, principal component analysis focuses on small differences in the asymmetry curves and works without any prior assumptions about the studied samples. We discovered that the PCA method works well in detecting phase transitions in muon spectroscopy experiments and can serve as an alternative to current analysis, especially if the physics of the studied material are not entirely known. Additionally, we found out that our ML technique seems to work best with large numbers of measurements, regardless of whether the algorithm takes data only for a single material or whether the analysis is performed simultaneously for many materials with different physical properties. 
\end{abstract}
\submitto{\JPCM}
\maketitle
\ioptwocol
\section{Introduction}
\par
Machine learning (ML) methods are now widely used in many areas of physics, usually as a tool to analyse large amounts of data \cite{Zdeborova:2017eo, Carleo:2019hc, Mehta:2019ju}. 
These techniques are particularly useful in regression, classification and dimensionality reduction tasks which are often required in processing scientific data. Specifically in condensed matter physics, ML is well suited for many tasks ranging from predicting materials properties based on existing databases and pattern recognition in specific experimental data to analysing theoretical models of quantum materials. Prominent examples include the prediction of novel materials \cite{Sumpter:2015ks,Xue:2016ev,Conduit:2017gu}, identification of phase transitions in models of magnetic materials starting from Ising models \cite{Wang:2016fu, 2017NatPh..13..431C, 2017NatPh..13..435V, Wetzel:2017ko, Broecker:2017kw, hu2017discovering}, reaching complex spin liquids in Heisenberg systems \cite{Greitemann:2019fu} and the detection of entanglement transitions from simulated neutron scattering data \cite{Twyman}. Machine learning algorithms were also proven to be state of the art techniques in simulations of wave functions \cite{Carleo:2017cn} or density matrices \cite{Nagy:2019bv, Vicentini:2019dn, Hartmann:2019hd, Yoshioka:2019ib} for many-body quantum systems and the tomographic reconstruction of many-body wave functions from experimental data \cite{Torlai:2018bd}. 

Much of the research in this area so far is concerned with simulation or analysing simulated data,
however it has also been shown that such techniques can detect phase transitions from piezoelectric relaxation measurements~\cite{Li2018Mar} or discovering existence of translational symmetry-breaking states from real, electronic quantum matter images \cite{zhang2019machine}.   
Here we want to apply a simple dimensionality reduction algorithm to real data from muon spin rotation ($\mu$SR) experiments~\cite{intro-muons} to see if we can detect phase transitions for a range of different materials. 
We decided to use the data from this type of experiment since models used in $\mu$SR data analysis require previous understanding of the local environment, which is not always easily available.
Therefore, as an alternative, we propose the use of linear principal component analysis (PCA), a simple unsupervised ML technique which does not make any prior assumption, yet is known to reveal correlations within the data. By demonstrating that this approach works, we propose that it may serve as a more unbiased way of detecting phase transitions observed in $\mu$SR experiments. In this paper we apply PCA to $\mu$SR data from a small number of superconducting and magnetic materials whose physics are known to differ widely from each other. In particular we explore the technique for data from time reversal symmetry breaking (TRSB) superconductors, which are among the most difficult to analyse, since changes in experimental data are very subtle. Other materials that we have tested are a symmetry breaking antiferromagnet (BaFe$_2$Se$_2$O) and a spin liquid (LuCuGaO$_4$). We find some evidence that PCA can detect important features such as phase transitions. We also find that when the system is trained on all the materials, taken together, the results improve -- even though the materials chosen have different underlying physics. 
\par
The paper is organised as follows. In section \ref{sec:KuboToyabe}, we briefly present the set up of the muon spectroscopy experiment and the current method of analysing the data from it. In section \ref{sec:PCA}, we present the principal component analysis in general and how we used it in practice. Then, in section \ref{sec:results}, we move on to results of applying PCA to data from different materials and discuss in detail how the method performs. We summarise the results in section \ref{sec:conclusions}.
\section{Muon spectroscopy experiment}
\label{sec:KuboToyabe}
\par
The general setup of a $\mu$SR experiment design to measure the local magnetic environment consists of spin-polarised muons being implanted into a sample, which is surrounded by multiple positron detectors. Once they enter the sample, muons will interact with the atoms causing them (muons) to thermalise and eventually implant themselves at some sites of the system. The spin of the muons will start to precess due to the local magnetic field and the muons will eventually decay into positrons and neutrinos with a mean life time of 2.2 $\mu$s. The positron velocity direction is directly connected to the muon spin orientation at the time of decay \cite{Schenck1985Jan,s_f_j_cox_implanted_1987,Blundell1999,Lee1999} and therefore the intrinsic magnetic field of the sample will affect the final distribution of positron detection events.
\par
\begin{figure}[t]
\begin{center}
    \includegraphics[width=0.99\linewidth]{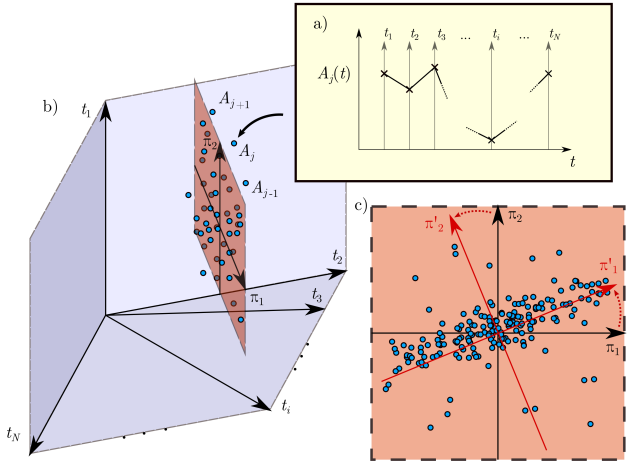}
\end{center}
\caption{Representation of the principal component analysis in high-dimensional data space. The asymmetry functions $A_j(t)$ consist of $N$ real values representing time windows $t_i$, $i=1,2,...,N$. Each of $t_i$ can be thought of as independent dimension (a). In this framework, we can represent each individual asymmetry function $A_j$ as a point in $N$-dimensional space (b)\footnotemark. We expect correlations between different asymmetry functions, which means that the data can be projected into a smaller subspace of initial $N$-dimensional space, without loss of information. The actual principal component analysis (c) can be represented as a rotation of initial coordinate space ($\pi_1$, $\pi_2$) into a new ($\pi'_1$, $\pi'_2$) so that most of the covariance is captured by the $\pi'_1$ dimension. The vectors in new basis ($\pi'_1$, $\pi'_2$) are called principal components and are usually numbered according to the amount of covariance they hold. Note: The projection onto $\pi_1 \times \pi_2$ plane is not a part of principal component analysis. The PCA rotates whole data space (after removing the average so that the cluster of data is centered at the beginning of coordinates) and then one can choose how many principal components (dimensions) must be used to represent the data well, based on the total covariance they hold.
}
\label{fig:1}
\end{figure}
A commonly used setup is to have symmetrical detectors in front of (F) and behind (B) the sample (with respect to the muon beam). The quantity that we are interested in is the difference in number of counting events between the two  detectors as a function of time $N_i(t)$, $i \in \{$F, B$\}$, called the asymmetry function
\begin{equation} \label{eq:1}
    A(t) = \frac{N_{\mathrm{B}}(t) - N_{\mathrm{F}}(t)}{N_{\mathrm{B}}(t) + N_{\mathrm{F}}(t)}.
\end{equation}
The analysis of the data involves fitting specific asymmetry curves to the experimentally-obtained curve. Given some knowledge of the underlying physics for a particular material and/or some justified assumptions, a model can be formulated, and the asymmetry curve can be derived from it. In some simple cases appropriate closed-form expressions can be derived~\cite{Kadono1989, Uemura1999}, though more generally {\it ad hoc} calculations are necessary \cite{deRenzi}. For some systems, our understanding is still not sufficiently developed for such predictions - for instance, the theory of zero-field muon spin relaxation (ZF-$\mu$SR) in superconductors with broken time-reversal symmetry (TRS) is still in its infancy~\cite{Atsushi2015Aug}.

In practice, for complex systems it is customary to use a phenomenological expression featuring several adjustable parameters. Electronic order can then manifest as a temperature-dependence of those parameters. For instance, in ZF-$\mu$SR investigations of superconductors~\cite{Ghosh2020Mar-Review} one often fits:
\begin{equation} \label{eq:2b}
    A_\mathrm{phen.}(t) =  A_0 G_\mathrm{KT}(\sigma, t) \exp(-\lambda t) + A_\mathrm{bckg}
\end{equation}
where $G_\mathrm{KT}(\sigma, t)$ is the Kubo-Toyabe function describing coupling to static, randomly-oriented magnetic moments~\cite{Hayano:1979_Zero_Field_muSR, Kubo:1981_StochasticTheory, Kadono1989, Uemura1999} with relaxation rate \(\lambda\) and Gaussian magnetic field strength distribution with standard deviation $\sigma$. The parameters $\sigma,\lambda,A_0,A_\mathrm{bckg}$ are then interpreted to describe distinct relaxation mechanisms. In conventional superconductors these parameters tend to evolve smoothly through the superconducting critical temperature, $T_c$. In other systems, marked changes in some of these parameters occur at $T_c$ \cite{Ghosh2020Mar}. These are often interpreted as evidence of broken TRS and in some systems this has been confirmed by Kerr effect or SQUID magnetometry. 
Quite frequently, it is found that only one of the fitting parameters in Eq.~(\ref{eq:2b}) depends on temperature. This is usually either $\sigma$ or $\lambda$, which naturally leads to a classification of TRS-breaking superconductors. We note, however, that the relaxation rates involved are very small, meaning that only a small portion of the curve described by Eq.~(\ref{eq:2b}) is represented in the experimental data sets (due to the finite lifetime of the muon). As a result, this classification may not always be as robust as would be desirable. For instance, some superconductors that are expected to have very similar underlying physics can fall in different classes. Such is the case of the proposed nonunitary triplet superconductors LaNiC$_2$~\cite{Hillier2009Mar} and LaNiGa$_2$~\cite{Hillier2012Aug}, whose asymmetry functions are best described by a temperature-dependent $\sigma$ and $\lambda$, respectively, in spite of experimental~\cite{chen_evidence_2013,weng_two-gap_2016} and theoretical~\cite{Ghosh2020Mar-Review} evidence of very similar underlying physical mechanisms. Likewise, the muon spin relaxation rate in spin glasses can often be described by a stretched exponential function (with temperature-dependent exponent), reflecting the variation in local spin fluctuation rates as well as non-exponential decay at muon sites ~\cite{Nuccio:2014fu,Campbell:1994kq,Keren:1996ec}. However, fitting experimental data can give parameter values that are not expected from standard models/numerical analysis ~\cite{Yadav:2019ez}. In conclusion, it would be highly desirable to have a way of analysing the temperature-dependence of $\mu$SR spectra that can detect electronic ordering transitions without the need to assume any {\it a priori} fitting functions.
\footnotetext{In panel b), each of the directions $t_1, t_2,..., t_N$ should be understood as being orthogonal to any other, thus spanning an $N$-dimensional target space representing a full dataset from an individual measured asymmetry function as a function of time.}

\begin{figure*}[t]
\begin{center}
    \includegraphics[width=1\linewidth]{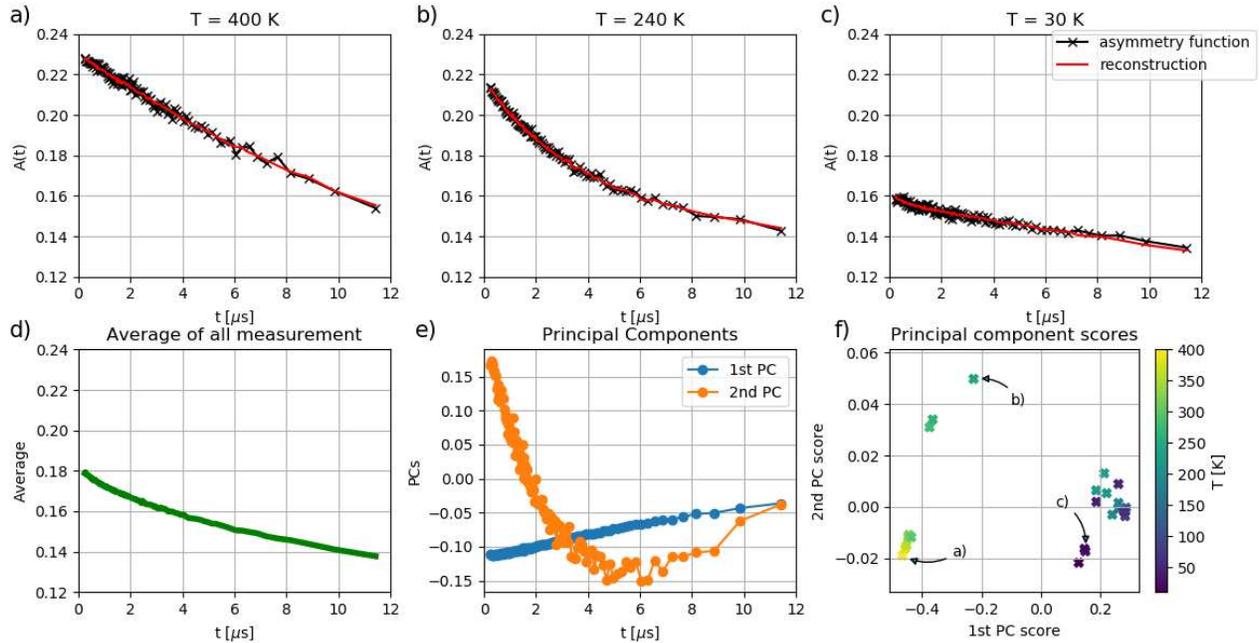}
\end{center}
\caption{An illustration of how principal component analysis can be used to reduce the dimensionality of a muon data set. The set consists of a sizeable number of  experimentally-obtained muon asymmetry functions $A(t)$. The black curves in panels (a-c) present three particular examples. Each curve has 110 time stamps and therefore constitutes a point in a 110-dimensional space. PCA yields a small number of  principal components (PCs) which, through linear combination, can accurately describe any curve in the data set. In our case, we find the two PCs shown in panel (e). The reconstruction of the original data using the PCs and the average (d) (see equation (\ref{eq:5})) can be obtained by the formula $\mbox{Reconstruction} = \mbox{Average} + \mbox{1st PC score} \times \mbox{1st PC} + \mbox{2nd PC score} \times \mbox{2nd PC}$. From that we can interpret the principal components as the most common deviations from the average curve. The reconstructions are shown, for our three examples, by the red curves in panels (a-c). This gives an accurate reconstruction and therefore enables us to represent each curve by a single point on a two-dimensional plane (f). For this example we used 25 $A(t)$ curves for the material BaFe$_2$Se$_2$O$_4$ obtained at 25 different temperatures.
}
\label{fig:2}
\end{figure*}
\section{Principal component analysis}
\label{sec:PCA}
To analyse the data from a muon spectroscopy experiment without making assumptions about the physical nature of the materials, we decided to use an unsupervised machine learning technique called principal component analysis (PCA)\cite{Wang:2016fu, geron2019, jolliffe2016principal}. The concept behind it -- in the context of muon spectroscopy experiment and asymmetry functions -- is presented in figure \ref{fig:1}. We can think about different experimental measurements as points in some data space with $N$ dimensions. In the case of muon spectroscopy, each dimension $i = 1, 2,...,N$ represents a time window $t_i$, within which the positron detections are measured. If the measurements are not random but correspond, for example, to the same material at different temperatures, we expect correlations between those points. PCA can detect these correlations by first removing the average of all experimental curves, then measuring the covariance for each dimension and linearly transforming the coordinates so that the new basis of the data space consists of only few directions that capture most of the covariance. The vectors of this new basis are called principal components (PCs) and can be thought of as the most common deviations from the average curve. We can reconstruct all of the measurements used in the analysis by adding to the average a linear combination of principal components. We can also represent each curve by specifying its projections onto the PCs, which are often called principal component ``scores''.  Thus, PCA provides us with a more compact description of the experimental data and additionally we can recover information about linear correlations from their magnitudes (or PC scores) and shapes (the principal components, or PC vectors). 
\par
In the example shown in figure \ref{fig:1} c), most of the data lies in two-dimensional space $\pi_1 \times \pi_2$. PCA finds new orthogonal directions ($\pi'_1$, $\pi'_2$), because there exist linear correlation between the $\pi_1$ and $\pi_2$ coordinates of data points. We can now specify each asymmetry curve by its projection onto $\pi'_1$, whereas before we would have to state both $\pi_1$ and $\pi_2$ coordinates. We do lose some information about the individual data points in this way, but we gain in the more compact representation of asymmetry curves. Usually, more than one principal component is needed to represent the data well. The number of important PCs varies with different data sets and can be decided by looking at how much covariance each principal component holds. 
\par
We present a more specific description of the PCA method. Each measurement can be represented as a vector $\mathbf{a}_j = \left(A_j(t_{1}), A_j(t_{2}), ..., A_j(t_{N})\right)^T$,\footnote{Here $A(t)$ stands for the measured quantity as defined in equation (\ref{eq:1})} with its values equal to the values of asymmetry function at specific times and the index $j = 1,2,...,M$ taken to label the distinct measured asymmetry curves that we want to analyse by the algorithm. We further assume that all measurements were recorded for the same set of $N$ measurement times $t_i$, taken relative to the time for implanting the muon into the material. We combine the vectors $\mathbf{a}_j$ in column form to construct a matrix $\mathbf{A}$
\begin{equation}
	\mathbf{A} = \left[\begin{array}{cccc}
	A_1(t_1) & A_2(t_1) & ... & A_M(t_1) \\
	A_1(t_2) & A_2(t_2) & ... & A_M(t_2) \\
	\vdots & \vdots & \ddots & \vdots \\
	A_1(t_N) & A_2(t_N) & ... & A_M(t_N).
	\end{array}\right]
\end{equation}
In the next step we remove the mean of each vector dimension (i.e., averaging over rows of the matrix) so that the whole data is centered around the coordinate origin, as shown in figure \ref{fig:1}. We end up with a matrix $\mathbf{X}$ with elements given by
\begin{equation} \label{eq:5}
[\mathbf{X}]_{ij} = A_j(t_i) - \frac{1}{M} \sum_{k=1}^M A_k(t_i). 
\end{equation}
The most common way for obtaining principal components is to perform a singular value decomposition of $\mathbf{X}$. To this end, we evaluate the covariance matrix
\begin{equation}
\mathbf{S} = \frac{1}{M - 1} \mathbf{X}\mathbf{X}^T,
\end{equation}
such that the eigenvectors of $\mathbf{S}$ are the principal components and the corresponding eigenvalues indicate the amount of covariance captured by the given PC. If we write the eigenvectors into a matrix $\mathbf{U}$, then a table of scores $\mathbf{C}$ for each measurement can be obtained by the matrix product
\begin{equation}
\mathbf{C} = \mathbf{U}^T \mathbf{X},
\end{equation}
and the full reconstruction of the initial experimental data is expressed as
\begin{equation}
\mathbf{R} = \mathbf{U} \mathbf{C}^T.
\end{equation}
\begin{figure*}[t!] \label{fig:7}
	\begin{center}
		\includegraphics[width=1\textwidth]{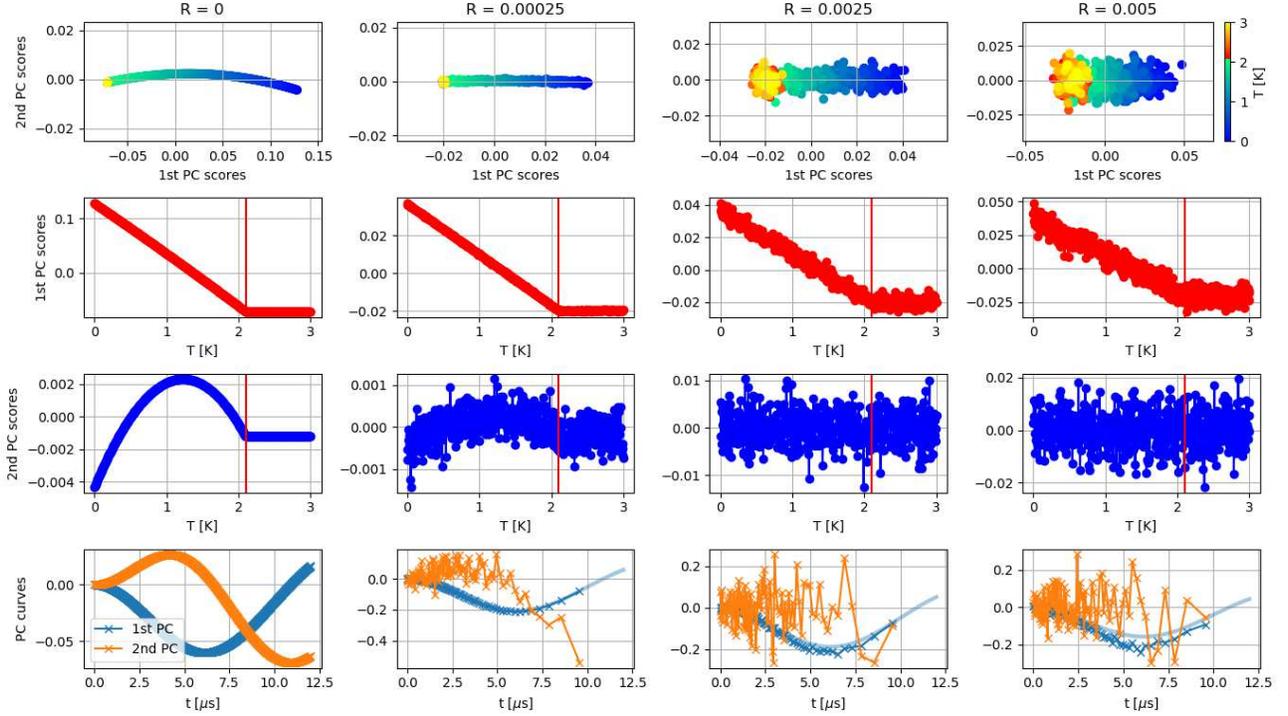}
	\end{center}
	\caption{Results of PCA performed on Kubo-Toyabe functions for a range of different simulated error. The third column ($R = 0.0025$) corresponds to error similar to our experimental measurements. On top row are the values of 1st vs 2nd PC scores and the change with temperature, 2nd and 3rd row are showing how PC scores change with temperature (the vertical red line corresponds to expected phase transition) and on bottom row the shapes of two most important principal component are shown. The scaled curve of first principal component without error was presented on the background of cases with noise.}
\end{figure*}  
\par
The previously discussed usefulness of the method derives from the fact that we can choose only the few PCs that capture most of the covariance in order to accurately reconstruct the initial data. Naturally, a large reduction in the number of relevant PCs does not have to arise for all possible data sets, as singular value decomposition only performs a linear transformation -- in particular, if the data has non-linear correlations the method will not perform well. Fortunately, looking at the eigenvalues of $\mathbf{S}$, one can decide if the linear PCA is sufficient, based on the decrease of PC scores which is often illustrated in a so-called scree plot of the PC scores against their index. 
\par
In order to account for the experimental noise in the data, we have re-binned raw data into new time windows according to the measurement error. Since the error increases with time, wider time windows are required at larger times to get comparable errors. Hence, available measurement points are more widely spaced at later times, as can be seen in figures \ref{fig:2} (a - c). It is important to re-bin all of the measurements simultaneously because all time-windows $t_1$, $t_2$,..., $t_N$ in our matrix $\mathbf{A}$ have to be the same for all columns for the PCA to be well defined. Note that this specification mirrors the treatment in regression methods, where less weight is attributed to data at long times to account for the larger measurement errors.
\subsection{Philosophy of our PCA approach}
\par
It is worth noting that in the PCA method presented above we do not have to make any assumptions on the shapes of asymmetry curves. There are no hyperparameters to vary, and SVD gives a unique representation of the sought asymmetry functions (up to a simultaneous change of sign of the principal components and the associated scores). Therefore we think that it provides an interesting alternative to fitting methods, where some initial knowledge of the probed material is needed. We would like to emphasize that it does not necessarily yield better results, but it can be applied to any type of input data reflecting all possible shapes of asymmetry functions. Furthermore, by examining scree plots of the PC scores, we are always able to judge how well the method performs in compressing the relevant data.

In figure \ref{fig:2}, we show an example that illustrates how the method detects changes in the shape of a set of experimentally measured asymmetry functions, obtained for a single material at different temperatures. The way in which those functions differ from each other is reflected in their respective scores for the 1st and 2nd principal components. Both high temperature (a) and low temperature (c) measurements have almost linear shape and they only differ in the values for the first principal component score. Looking at the 1st PC shape (panel (e), blue curve), we can see that it is also almost linear and when  
multiplied by a large negative value -- as it is for high temperature asymmetry function -- and then added to the average (d), it increases the overall slope. For the low temperature curve, it is added with positive sign, which means that it will instead decrease the slope. We can see that it is exactly the difference between two asymmetry functions (a and c). On the other hand the middle curve (b) differs mostly in second principal component scores from the other two. When the second PC vector (e, orange curve) is multiplied by a positive value and added to the average, it creates a more convex curve, which is reflected by the shape of the corresponding asymmetry function (b).
\begin{figure*}[t!]
    \centering
    \includegraphics[width=1\linewidth]{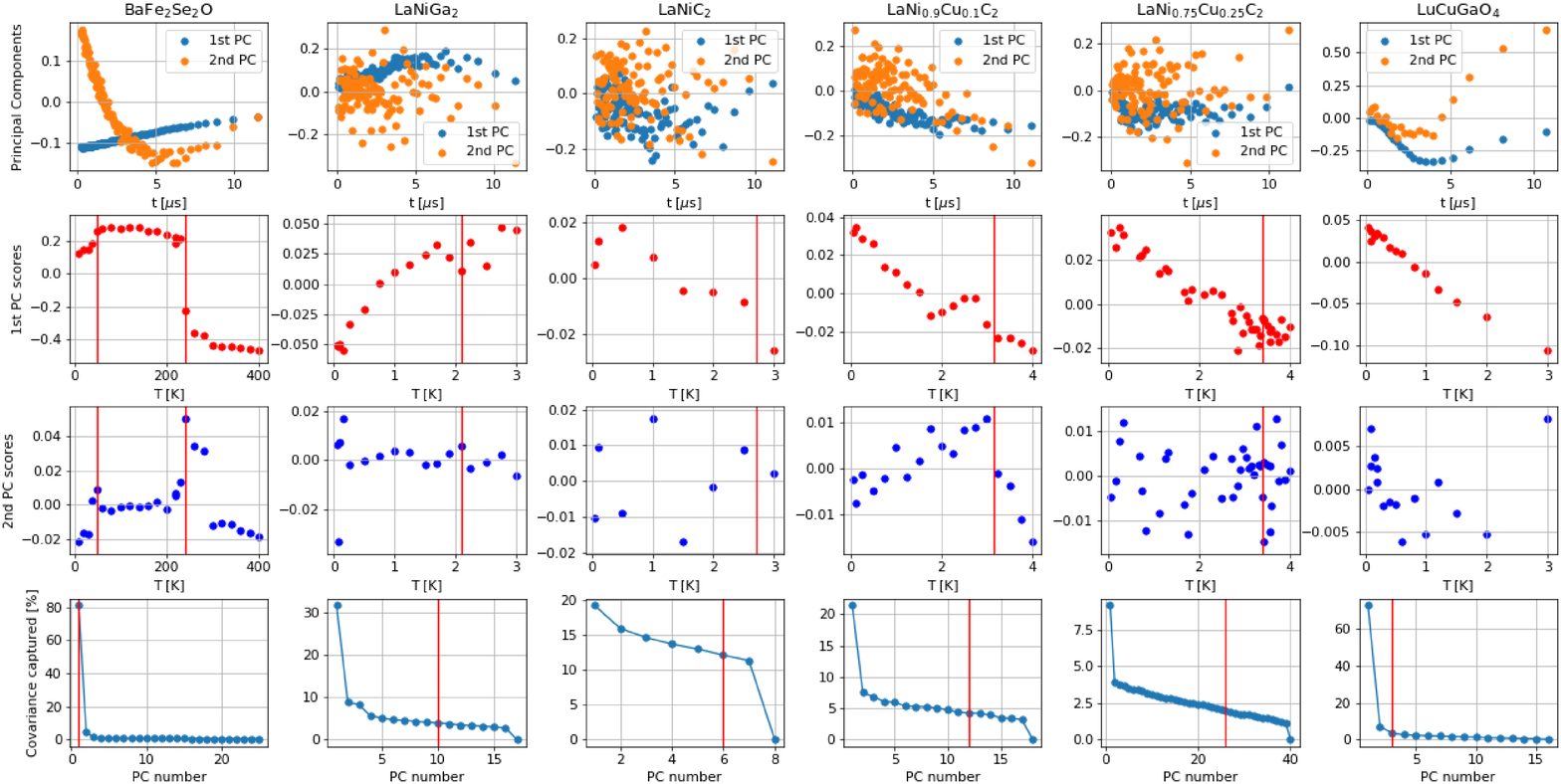}
    \caption{Results of principal component analysis performed independently for each material. The top row presents the shapes of the 1st and 2nd principal components. For almost all the cases second principal does not look as smooth as the 1st PC. The second and third row display the dependence of PC scores on  temperature for the 1st and 2nd PCs, respectively. The red vertical line indicate approximately where we suspect phase transitions to occur \cite{SC_criticaltemperature, Coles:2019ey}. The last row presents a scree plot for the amount of covariance that each principal components captures. Here, red lines indicates how many principal components are needed to capture 80 \% of the total covariance.}
    \label{fig:3}
\end{figure*}
\begin{figure*}[t!]
    \centering
    \includegraphics[width=1\linewidth]{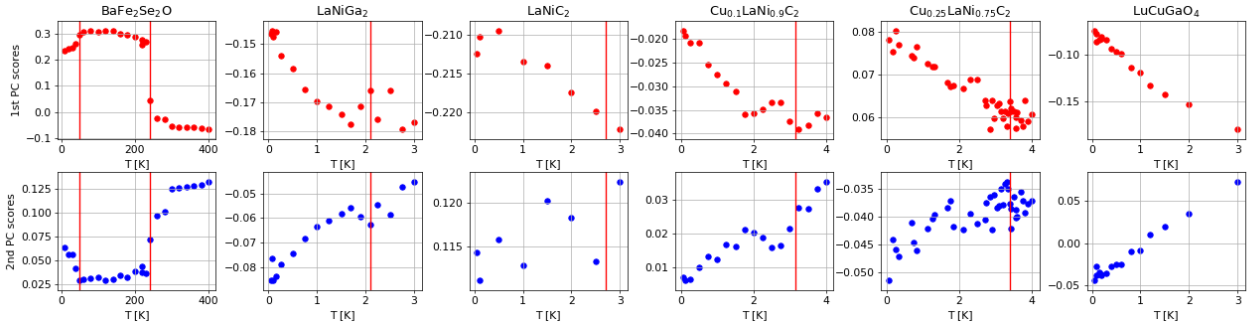}
    \caption{Results of principal component analysis performed simultaneously on experimental data from all materials. The first and second row display the dependence of PC scores on temperature for the 1st and 2nd PCs, respectively. The red vertical line indicate approximately where we suspect phase transitions to occur \cite{SC_criticaltemperature, Coles:2019ey}}
    \label{fig:4}
\end{figure*}
\begin{figure*}[t!]
    \centering
    \includegraphics[width=1\linewidth]{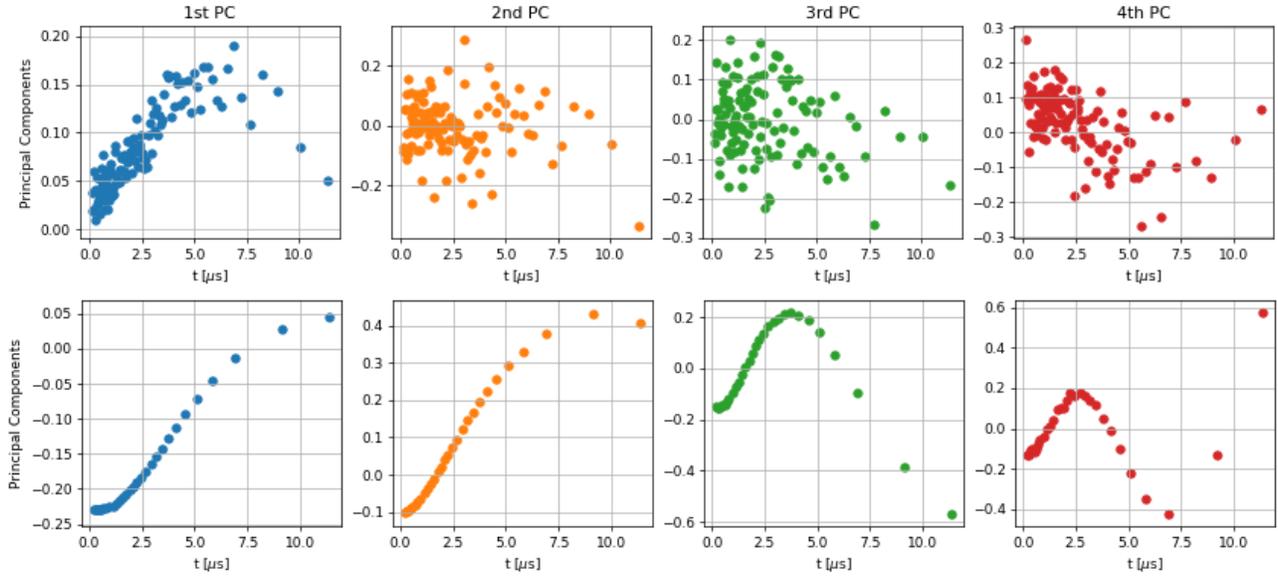}
    \caption{Comparison of principal component vectors as a function of whether the analysis is performed independently on the experimental data for each individual material, or jointly for all materials. The top row corresponds to PCs from the PCA of LaNiGa$_2$ exclusively, and the bottom row represent the same for the PCA of all materials processed jointly. One can see that PCs become much smoother functions after providing the algorithm with a larger amount of data.}
    \label{fig:5}
\end{figure*}
\begin{figure}[t!]
    \centering
    \includegraphics[width=1\linewidth]{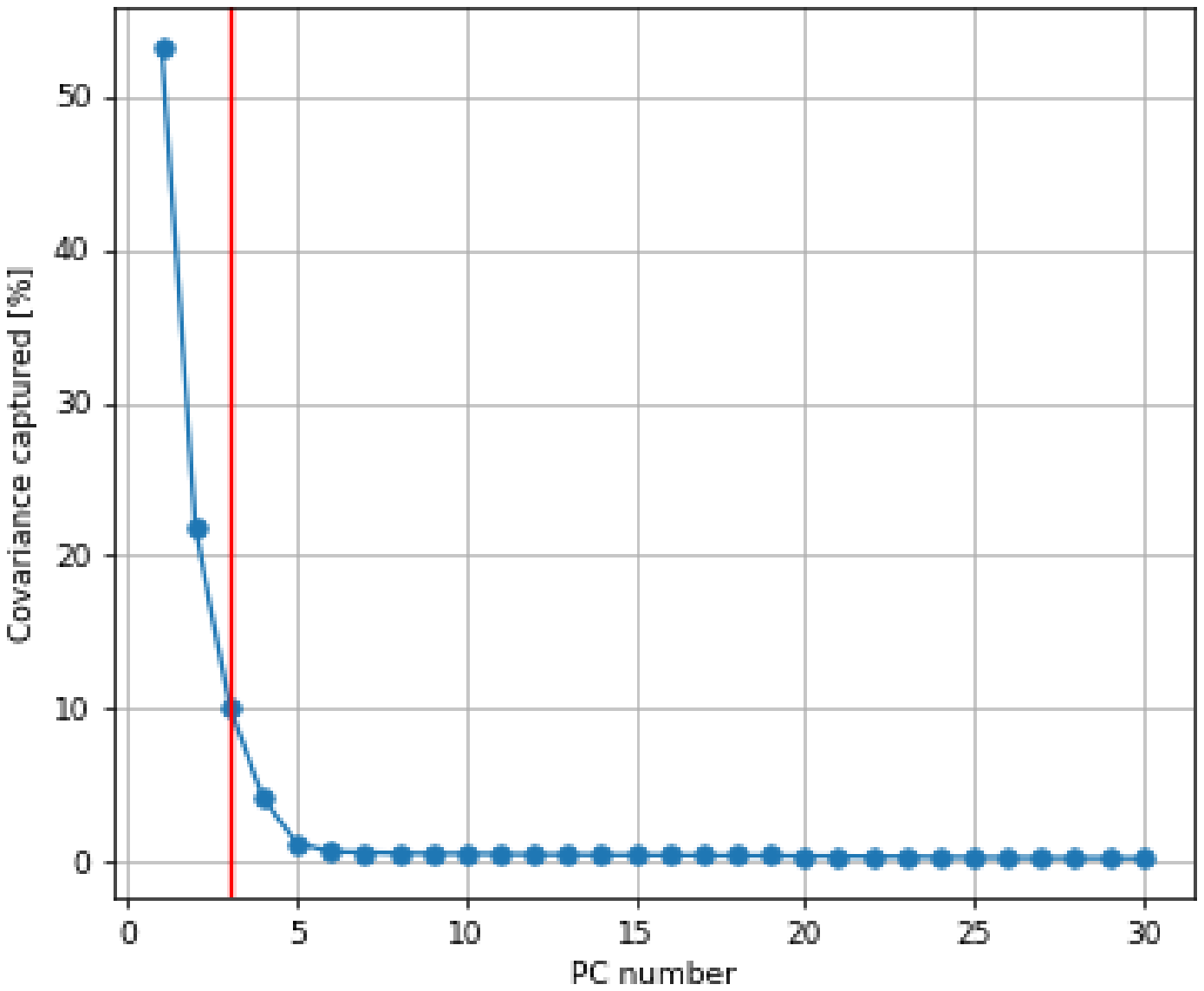}
    \caption{Amount of covariance captured by the principal components from the PCA of all the materials analysed simultaneously. 80 \% of the covariance is captured by the two first PCs.}
    \label{fig:6}
\end{figure}
\section{Results and discussion}
\label{sec:results}
\subsection{PCA for simulated data}
To illustrate characteristic results of performing PCA on asymmetry functions, we first consider an example application to synthetic data generated from model Kubo-Toyabe functions $G_\mathrm{KT}(\sigma, t)$ with added error $E(t)$. Each such simulated asymmetry function was taken from the general form given by
\begin{equation}
	A_\mathrm{sim}(t;T) = A_0 G_\mathrm{KT}(\sigma(T), t) \exp(-\Lambda t) + A_\mathrm{bckg} + E(t),
\end{equation}
where we have further encoded a dependency of $\sigma(T)$ associated with a symmetry breaking phase transition in many TRSCs, such that $\sigma(T)\sim const$ for $T>T_c$, while varying linearly below $T_c$.
The error values were generated from a Gaussian distribution $N(\mu=0, \Sigma_\mathrm{sim})$ centered on zero and with a standard deviation\footnote{We use the symbol $\Sigma$ for the standard deviation of simulated errors, to distinguish it from the parameter $\sigma$ of the Kubo-Toyabe form.} $\Sigma_\mathrm{sim}(t)$ depending on time after muon implantation as:
\begin{equation}
	\Sigma_\mathrm{sim} (t) = R (a^t + b).
\end{equation}
\par
Errors observed in real measurements increase with time $t$ due to the overall smaller number of events detected at later times. The parameters $A_0$, $\sigma(T)$, $\Lambda$, $A_\mathrm{bckg}$, $R$, $b$, and $a$ were chosen to match experimental data of one of the superconductor studied (LaNiGa$_2$). In addition to the parameters reflecting experimental conditions, we studied the effect of different error amplitudes $R$ (which in experiments would correspond to experiments undertaken with different amounts of time allocated for integrating the signal) in order to verify robustness of the PCA approach.
Our results from the application of PCA to this simulated data is displayed in figure \ref{fig:7}. We have included four possible cases of ``noise'' amplitudes ranging between no error and twice the error we expect from our measurements. 
The PCA on clean data clearly captures the transition temperature $T_c$ assumed in the simulated data, which separates regions of temperature with or without variation of the principal component scores with $T$.
The first principal score dependency is found to be very robust to added noise, even for the cases where the error is much larger than expected experimentally. By contrast, the second principal component does not seem to hold any useful information for realistic noise level. Note also the small overall scale of the second PC score. Nevertheless, the phase transition is always clearly visible in the 1st PC, which motivates using PCA for experimental data.

\subsection{PCA for experimental data}
We applied principal component analysis to data from zero-field muon spin rotation experiments for a range of different materials. Among them are time-reversal symmetry breaking superconductors (LaNiGa$_2$, LaNiC$_2$, LaNi$_{1-x}$Cu$_x$C$_2$), spin liquid (LuCuGaO$_4$) and an antiferromagnet (iron oxyselenide BaFe$_2$Se$_2$O). We first performed the analysis for each material separately. The shape of the two most important PCs and the dependence of the scores on temperature are presented on figure \ref{fig:3}. 
\par
Our technique worked best for the antiferromagnetic material (first column on figure \ref{fig:3}), for which both expected phase transitions are clearly visible. Although the magnetic behaviour of the antiferromagnet BaFe$_2$Se$_2$O is relatively simple  ~\cite{Coles:2019ey}, this understanding has been challenging to arrive at: a) $T_{\mathrm{N}}$ $\sim$ 240 K is clear from neutron powder diffraction experiments but is more subtle in magnetic susceptibility measurements ~\cite{Coles:2019ey,Lei:2012gc,Han:2012dy} due to the layered nature of the material; b) Magnetic susceptibility data collected on several samples suggest a magnetic phase transition at $\sim$ 115 K ~\cite{Coles:2019ey,Lei:2012gc} which is now thought to be due to Fe$_3$O$_4$--related impurities and is not intrinsic to the main phase ~\cite{Coles:2019ey}; c) There’s no evidence for the low-temperature $\sim$40 K phase transition from neutron powder diffraction ~\cite{Coles:2019ey} or heat capacity data ~\cite{Han:2012dy} and this phase transition is thought to involve freezing of spin fluctuations. It is striking that this unsupervised machine learning analysis correctly identified the two phase transitions intrinsic to BaFe$_2$Se$_2$O without the need for complementary data. We think that this reflects the strength of both the PCA analysis and the muon spin rotation technique.
\par
The changes in the asymmetry function are more subtle for the superconducting materials (second-fifth column on figure \ref{fig:3}), but the behaviours of PC scores still change at expected critical points. In the case of LaNiC$_2$ (third column), for which we only have one point above phase transition and therefore we do not expect visible change. That is confirmed in PC score plots. Worth mentioning is also the LaNi$_{0.9}$Cu$_{0.1}$C$_2$ case, in which there seem to be more than one critical point, at least in the behaviour of 1st PC score. That might be caused by some other phase transition but more probably it is caused by limitation of the method. One solution to that problem would be to look also at the 2nd PC score, where only one transition point is prominent. Overall, linear PCA seems to be performing better for the spin liquid and antiferromagnetic materials than for the time-reversal symmetry breaking superconductors analysed in this paper, as is evidenced in our scree plots (the last row on figure \ref{fig:3}). For the first four materials, even the last few PCs hold a significant amount of covariance\footnote{The singular value decomposition yields $\mbox{min}(N,M)$ singular values, so for the case of $N > M$ the covariance captured by the last PC will formally vanish, by definition.}. That may imply that the data has non-linear correlations or that we did not have not enough data available for these types of materials, since most ML algorithms perform better the more data is provided. It is important to note that we can still resolve the changes in the scores of 1st and 2nd PCs, at least for LaNiGa$_2$ and LaNi$_{0.9}$Cu$_{0.1}$C$_2$.
\par
Last studied material is a proposed spin liquid -- LuCuGaO$_4$. Muons have been used as a proof of a spin liquid state, as it can be argued that the resultant dynamics could show a plateau in the relaxation rate with reducing temperature where no long range order is detected \cite{KhuntiaSpinLiquid}. In our case the PCA shows no evidence for a phase transition, even though a plateau is observed, likely indicating there is no phase transition as the proposed liquid state is entered. 
\par 
Because the most significant principal components for the time-reversal symmetry breaking superconductors look similar for all the cases studied, in hope of improving results for TRSB systems, we proceed to apply PCA to all of the experimental data simultaneously. The results are presented in figures \ref{fig:5} and \ref{fig:6} and the comparison of principal components obtained from these two different PCA analyses is shown in figure \ref{fig:4}. The principal components are now much smoother functions and additionally, only three PCs are sufficient to capture 80\% of the observed covariance. The scores of the first principal component did not change much for all materials, despite their different physical properties. This is probably connected to the fact that all data come from the same type of experiment and all asymmetry functions are similar in general.

\section{Conclusions}
\label{sec:conclusions}

We have proposed the use of principal component analysis to process muon spin spectroscopy data, and in particular to aid with the identification of features relating to phase transitions in the probed materials. Our results demonstrate that the representation of the observed asymmetry functions in the space of principal component vectors is 
sensitive to changes in the physics of the observed system. In particular, the evolution of principal component scores as a function of tuning parameters provides 
insights into the location of possible phase transitions. Comparing this analysis to a more conventional approach, based on regression analysis using standard fitting functions, we find that PCA is typically at least as sensitive, if not more. 
More importantly, 
the PCA approach is free from any underlying assumptions about the physics of the observed material: rather than assuming a specific form of a fitting function (e.g.~Kubo-Toyabe or stretched exponential), PCA discovers the principal components that describe a given system, without human intervention. This is the salient feature of the method we put forward and it means that the same, universal analysis can be applied to any material. In addition we have found that the quality of the results is enhanced when data for multiple materials are analyzed as a joint dataset, even when the underlying physics of each system being considered are quite different. The ability to thus enhance understanding gained from a new experiment based on existing data goes beyond the possibilities of preexisting approaches, where data for each material is necessarily analyzed and fitted in isolation, and overarching commonalities are anticipated in advance by the formulation of a suitable fitting function. We anticipate this could offer great advantage when deployed in large-throughput user facilities. In particular, given the advantages gained from combining multiple data sets, our results suggest a new way to leverage recently-developed open-data tools and policies~\cite{ICAT,PANDATA}.
\par
We hope that our unsupervised ML approach to muon spectroscopy data analysis could become one of the standard tools used in that field. 
In addition to its virtue, noted above, of providing a
unified way of treating all muons data, 
we believe our approach can also accelerate future experiments, as the treatment within this framework will require less data to be collected before signatures of the physics can emerge---especially 
if data from previous experiments is used to enhance the analysis of new materials as outlined above. In addition, the simplicity of the analysis means that it could easily be performed immediately while experimental measurements are being taken, thus opening the possibility to inform the conduct of the experiment in real time. At the other end of the spectrum, it is also possible to conduct experiments where much larger data sets are gathered~\cite{Wilkinson2020Aug}. Our simulations suggest that our method applied to such data might yield valuable new insights into phase transitions. They also would be ideal additions to such past-experiment data bank. Given the advantages gained from combining multiple data sets, our results should encourage the community to gather historic and future measurements in a common database in order to harvest the benefits of this approach.

\section{Acknowledgments}
We would like to thank Stephen Blundell, Tom Lancaster and Roberto De Renzi for helpful discussion about the content of the paper.
\par
TT is supported by the EPSRC via a DTA studentship under grant no.~EP/R513246/1 and by the School of Physical Sciences, University of Kent. SR and EEM are grateful to Mr Ben Coles (for BaFe$_{2}$Se$_{2}$O synthesis) and to Dr Fiona Coomer (experimental support) for the $\mu$SR data from reference ~\cite{Coles:2019ey}.
JQ acknowledges support from the EPSRC under the project  “Unconventional superconductors: New paradigms for new materials” (Grant No. EP/P00749X/1). GM gratefully acknowledges support by the Royal Society under University Research Fellowship URF\textbackslash R\textbackslash 180004.

\section{Individual author contributions}
T.~Tula implemented the PCA algorithm and performed the analysis of the simulated and experimental data presented in this paper under supervision from J.~T.~Quintanilla and G.~M\"oller. S.R.~Giblin, A.~D.~Hillier, E.~E.~McCabe and S.~Ramos provided, formatted and commented on the experimental data. D.~S.~Barker performed a preliminary study of PCA applied to experimental and simulated $\mu$SR data under supervision of J.~Quintanilla, with further input from S.~Gibson. T.~Tula wrote the manuscript in close consultation with G.~M\"oller and J.~Quintanilla and with input from all co-authors.
\section*{References}
\bibliographystyle{unsrt}
\bibliography{bibliography}
\end{document}